\documentclass[aps,twocolumn,pre,showpacs,superscriptaddress,amsmath,amssymb]{revtex4-1}
\usepackage{amsmath,comment}
\usepackage{amssymb,graphicx}
\usepackage{color}

\usepackage{fourier}

\begin{document}
\title{Cost-efficient vaccination protocols for network epidemiology}

\author{Petter Holme}
\affiliation{Institute of Innovative Research, Tokyo Institute of Technology, Tokyo, Japan}
\author{Nelly Litvak}
\affiliation{Department of Applied Mathematics, University of Twente, Enschede, Netherlands}

\begin{abstract}
We investigate methods to vaccinate contact networks---i.e.\ removing nodes in such a way that disease spreading is hindered as much as possible---with respect to their cost-efficiency. Any real implementation of such protocols would come with costs related both to the vaccination itself, and gathering of information about the network. Disregarding this, we argue, would lead to erroneous evaluation of vaccination protocols. We use the susceptible-infected-recovered model---the generic model for diseases making patients immune upon recovery---as our disease-spreading scenario, and analyze outbreaks on both empirical and model networks. For different relative costs, different protocols dominate. For high vaccination costs and low costs of gathering information, the so-called acquaintance vaccination is the most cost efficient. For other parameter values, protocols designed for query-efficient identification of the network's largest degrees are most efficient.
\end{abstract}

\maketitle

\section{Introduction}

Infectious disease is a major burden to global health. Infections spread from person to person over human contact networks. The propagation speed is an emergent property of both the pathogenesis in the infected individual and the contacts between people. By understanding the contact networks, we should thus be able to better predict and mitigate disease outbreaks. These are the premises of network epidemiology~\cite{keeling_rev,ps_rev}---one of its most active questions being how to exploit the contact network in targeted vaccination campaigns~\cite{lv,wang_etal_statphys}. Until now, targeted vaccination has mostly been a theoretical topic. The medical practice of network-based immunization has been very limited to both few cases and simple methods---the most famous being ``ring vaccination''~\cite{gies}. This strategy was used to eradicate smallpox and works by vaccinating all network neighbors of an infectious person~\cite{ring}. Nevertheless, network immunization could be important in future disease control, especially for sexually transmitted infections (where the network links are evident)~\cite{lea:sex} or livestock diseases (where one node is a farm and links are connections by transport)~\cite{colizza}.

In the theoretical literature, the problem of targeted vaccination has typically been formulated as follows. Given some knowledge of the contact network, identify the individuals that are potentially most important for disease spreading. To carry out a targeted vaccination campaign, one would first need to gather information about the contact network, then use this information to vaccinate (or otherwise reduce the impact of the important individuals). There are thus three major costs involved in such an endeavor: the cost of the disease itself (that we can use as our base unit), the cost of gathering the information about the network $c_\mathrm{info}$ (in units of the cost of a person getting the disease) and the cost of vaccinating $c_\mathrm{vacc}$. We can thus evaluate the cost efficiency of a vaccination protocol by measuring the net saving $\chi$ per person in units of the cost of sick individuals
\begin{equation}
\label{eq:chi}
N\chi(f) = \Omega-\Omega'(f)-Nfc_\mathrm{vacc} - n(f) c_\mathrm{info}
\end{equation}
where $\Omega$ and $\Omega'$ are the expected outbreak sizes (number of individual who had the disease after it became extinct) respectively without and with using vaccinations, $N$ is the number of individuals, $f$ is the fraction of individuals to vaccinate and $n$ is the number of inquiries needed to obtain information. { Obviously, $\Omega$ correspond to the no-vaccination scenario and thus does not depend on $f$.} Both $n$ and $\Omega'$ depend on the specific vaccination protocol, but we drop this information in Eq.~\eqref{eq:chi} for brevity.

By reformulating the vaccination problem as a cost-optimization problem, one can evaluate the protocols proposed in the literature in a way more useful for decision makers. In this paper, we use this approach to evaluate eight vaccination protocols for many kinds of cost scenarios and underlying networks. We use eight different empirical networks of human contacts (representing sexual interaction or proximity). We also use the configuration model---a popular method to generate synthetic uncorrelated random networks given a degree sequence.

Before proceeding to the details of our approach we will give a brief overview of the recent analytical advances on the vaccination problem. The simplest vaccination protocol is just to vaccinate random individuals---the {\it Random (R)} protocol---which often serves as a baseline in the literature, see e.g.\ Refs.~\cite{nv,Britton2007graphs,Lelarge2009efficient_vacc,Ventresca2013vacc_strategies}. In a seminal paper, Cohen, Havlin and ben Avraham~\cite{nv} proposed the more effective {\it Acquaintance (A)} vaccination. In their approach, one also starts by randomly selected individuals, but does not vaccinate these, rather, asks them to name someone they met (in such a way that contagion could occur). In an uncorrelated network, the probability of meeting a node of degree $k$ in such an approach, is proportional to $k$. It is important to vaccinate high-degree nodes, not only because they have more people to spread the disease to, but also more people to get the disease from. 

Let $f_c$ denote the fraction of population that must be vaccinated in order to prevent a global outbreak. Formally, as $N\to\infty$, $f_c=\inf\{f: \Omega'(f)/N=o(1)\}$, and we will use a superscript for $f_c$ to denote a specific vaccination protocol. It was shown numerically in Refs.~\cite{nv,Britton2007graphs} that $f^A_c<f_c^R$. An implicit analytical expression for $f_c^A$ in uncorrelated networks (configuration model) was derived in Ref.~\cite{Britton2007graphs}. Similar results were obtained in Ref.~\cite{Lelarge2009efficient_vacc} for a more general model of infection spreading, in Ref.~\cite{ball2013acquaintance} for imperfect vaccine, and in Ref.~\cite{Deijfen2011epidemics} for the weighted configuration model, where weights of the edges represent contact probabilities.

A large empirical study based on the 2006 census of the Greater Toronto Area~\cite{Ventresca2013vacc_strategies} suggests that vaccination of top-degree nodes---the {\it Degree (D)} vaccination protocol---is most effective. However, such strategy requires information about the entire network, which makes it hard to implement. For analytical results on degree-based vaccination and an implicit expression for $f_c^{\it D}$, we refer to  Ref.~\cite{Lelarge2009efficient_vacc}. In this paper by optimizing Eq.~(\ref{eq:chi}) rather that $f_c$, we confirm that the {\it Degree} protocol is never the most efficient one: in all scenarios, the cost of the complete knowledge does not justify the gain in $\Omega'$. 

In addition to the {\it Acquaintance} protocol, we consider two strategies, recently developed for quick detection of high-degree nodes: the {\it Random walk (RW)} strategy~\cite{nelly}, and the {\it Two-step heuristic} ({\it TSH})~\cite{twitter}.  We also consider two other protocols that require complete knowledge of the network---{\it Coreness} and {\it Collective influence} ({\it CI}). See  below for a complete description of all protocols.

\section{Methods}

In this section we introduce the methods, data sets and network models we use.

\subsection{SIR simulation}

We assume that an infectious disease is spreading over a static contact network represented as a graph $G=(V,E)$. $V$ is the set of $N$ vertices, or nodes, representing individuals; $E$ is the set of $M$ undirected edges representing pairs of individuals between whom the disease may spread. The nodes are, at any given time, in one of three states---susceptible (S), infectious (I) or recovered (R). Susceptible nodes do not have the disease, but can get it. Infectious nodes have the disease and they can spread it. Recovered nodes do not have the disease and cannot get it. We assume a disease outbreak starts at time $t=0$. At the beginning all nodes are susceptible, except a randomly chosen node that is infectious. If an edge is between one susceptible individual and one infected individual, then the susceptible becomes infectious at rate $\beta$. Every infectious recovers at rate $\nu$. In this setting, the infection and recovery times are independent exponential random variables, and an infectious node transfers a disease through an edge before getting recovered with probability $\beta/(\beta+\nu)$.

The SIR model is essentially determined by the ratio between $\beta$ and $\nu$. In the well-mixed, differential equation version of the SIR model, this ratio is called $R_0$. The actual values of $\beta$ and $\nu$ are only needed to calculate the time to reach the peak prevalence, extinction etc. In this paper, we set $\nu=1$ which is equivalent to saying that we are measuring the time in units of $1/\nu$. In order to simulate this model, it is efficient to perform one infection or recovery event every iteration of the algorithm. The probability of the next event being an infection is
\begin{equation}\label{eq:prob}
 \frac{\beta M_{\rm SI}}{\beta M_{\rm SI} + N_{\rm I}},
\end{equation}
where $M_{\rm SI}$ is the number of edges between infectious and susceptible individuals, and $N_{\rm I}$ is the prevalence (number of infectious individuals~\cite{holme_versions}). The time increment since the last iteration is, on average, $1/(\beta M_{\rm SI} + N_{\rm I})$. Thus, to record time (in units of $1/\nu$), one adds this amount every iteration to a variable representing time, which results in a discretized version of the system. (We note that since our analysis is based only on average characteristics of the system, such discretization does not affect the results.) If an infection event is not performed, one performs a recovery event. In an infection event, the S-I edge is chosen randomly among all S-I links. Similarly, in case of a recovery event, the infectious individual (to recover) is selected uniformly at random among all infectious individuals.

For all contact networks and parameter values, we use $300,000$ or more runs of the SIR model for averages.  Since each run represents an independent realization of the (random) costs of the entire process, we used the normal approximation of the sample average,  to verify with 99\% confidence intervals that our evaluations of mean costs were very accurate. The exact values of the confidence intervals are not informative for our purpose and thus are omitted.  We use $\beta=1/32,1/16,1/8,1/4,1/2,1,2,4,8,16,32$ and (as mentioned) $\nu=1$.

\subsection{Vaccination protocols}

We compare the performance of seven vaccination protocols---five of these have been analyzed in the literature, and two are proposed by us in this work (but derived from a cost-efficient way of finding the highest degree vertices). The vaccination protocols range from simple to complex and use different amount of information about the network.

\subsubsection{Random vaccination}

The simplest way of vaccinating a fraction $f$ of a population is to just pick $fN$ persons uniformly at random~\cite{ps_vesp}. In this case, we can assume the information cost to be zero as all we need is a list of contact information of the population.

\subsubsection{Acquaintance vaccination}

An elegant way of exploiting the network structure to find high-degree individuals to vaccinate is the \textit{Acquaintance} vaccination scheme by Cohen, Havlin and ben Avraham~\cite{nv}. In the literature it is often assumed that each individual is sampled a Poisson($f$) distributed number of times, and each time a sampled individual names one neighbor to vaccinate. When the neighbor has already been vaccinated, no vaccination occurs and the next individual is sampled randomly. Then, the average fraction of vaccinated individuals $v(f)$ is smaller than $f$. The exact formula for $v(f)$ is given e.g.\ in Refs.~\cite{Britton2007graphs,Lelarge2009efficient_vacc}. Naturally, $v(f)$ is close to $f$ when $f$ is small. Here we assume that when a randomly sampled individual names a contact, which has already been vaccinated, then the individual is asked to name another contact. We discard the rare cases when all contacts of a random individual have already been vaccinated, and thus assume that $v(f)=f$. Then the information cost of this protocol is $fNc_\mathrm{info}$, since one needs to make an inquiry to one node for every node that is vaccinated. 

\subsubsection{Random-walk vaccination}

If one is willing to spend more effort on mapping out the network, one can do significantly better than the acquaintance vaccination in finding the high-degree vertices. This is the idea of \textit{Random walk} vaccination. Under this heuristics one keeps a list of the $fN$ vertices with highest observed degree that is updated during a random walk of inquiries. This is based on Ref.~\cite{nelly} that proposed this method to find high-degree nodes in the World Wide Web  and social networks in a cost-efficient way.  Let $k_i$ be the degree of node $i$. When the random walk is at node $i$, it jumps to a random node with probability $\alpha/(k_i+\alpha)$, and with complementary probability it proceeds to a randomly chosen neighbor of $i$. The rationale is that the stationary probability of $i$ in such random walk is proportional to $k_i+\alpha$. We use $\alpha=3$ following the recommendation from Ref.~\cite{nelly} that $\alpha$ should be of the order of the average degree. This value will be the same in all networks because the performance is robust with respect to $\alpha$. The second parameter $m$ is the number of steps in the random walk. Rather than fixing this parameter, we will use the value that optimizes $\chi$.

The cost of this protocol is the number of steps the random walk continues to a neighbor of the present node (rather than jumping to a random node) times $c_\mathrm{info}$. (On average, in stationarity, the information costs are $mc_\mathrm{info}(1-\alpha/(\langle k\rangle +\alpha))$, cf.\ Ref.~\cite{nelly}.

\subsubsection{Two-step heuristic}

We also try a protocol that, like the \textit{Random walk} in the previous section, was developed to cost-efficiently identify high-degree nodes in social media. We call it the \textit{Two-step heuristic (TSH)}. Just like \textit{Random walk} it has a parameter to tune the amount of information used in the search process~\cite{twitter}. This protocol consist of two stages. In the first stage, one randomly chooses $n_1$ nodes and considers a reduced network of these $n_1$ nodes and their neighbors. In the second stage one measures the exact degrees of the $n_2$ highest-degree nodes of the reduced network. For simplicity, we set $n_1=n_2=n$ (which is not far off the expected optimal parameter setting~\cite{twitter}). This gives $n(f,G)=2n$, and the total information cost $2nc_\mathrm{info}$. 

\subsubsection{Degree}

Since both \textit{Random walk} and \textit{TSH} aim at being cost-efficient methods to rank nodes according to their degree, we also use the correct values of the degree (which could only be obtained by knowing the entire network). The information cost of this protocol is thus $Nc_\mathrm{info}$. This has also been discussed as a vaccination protocol~\cite{ps_vesp}.

\subsubsection{Coreness}

There are other structures than degree that could be exploited for mitigating disease spreading. Coreness captures, not only the degree of a node, but also increases with the connectedness of a node's neighborhood. The idea that dense clusters (``core groups'' in the epidemiological literature) are important for disease spreading dates back to Ref.~\cite{coregroup}. Coreness is not the only metric to capture this property, but a simple and straightforward one. It is the byproduct of a $k$-\textit{core decomposition}, which is a way to analyze the network by successively removing nodes from it. Specifically, at level $k$, one deletes all nodes with degree $\leq k$. If nodes get degree $\leq k$ during the deletion process, one deletes these too, until all nodes have degrees larger than $k$. The coreness value of a node is the $k$-value when it was deleted.

The coreness as an estimate of importance with respect to disease spreading was proposed by Ref.~\cite{core}, and further refined in Ref.~\cite{hebert2016multi}. To use it, one would need to map out the entire network, i.e.\ all its $M$ edges.  However, in reality, the inquiries will be implemented node by node. Therefore, we choose  a simplified approach, in which  we assume that knowing the complete network takes one inquiry per node, i.e.\ the total information cost is $Nc_\mathrm{info}$. Note that this is a more demanding inquiry, because it requires an individual to list all its neighbors.  Still, we use the same cost, meaning the performance of {\it Coreness} relative to its cost will be slightly  overestimated compared to the above protocols.

\subsubsection{Collective influence}

Finally, we use a yet more elaborate algorithm that, like coreness, requires full information about the network. We stick with the author's rather non-descriptive name \textit{Collective Influence} (\textit{CI})~\cite{ci}. It starts by defining a quantity
\begin{equation}
x_l(i) = (k_i-1)\sum_{j:d(i,j)=l}(k_j-1)
\end{equation}
where $k_i$ is the degree (number of neighbors) of $i$, $d(i,j)$ is the distance (fewest number of edges in any path) between $i$ and $j$. The algorithm proceeds by deleting the node of largest $x_l(i)$, then recalculating $x_l$ for the reduced network and repeating the procedure until $fN$ nodes are deleted. As $l$ grows, the ranking stabilizes but the computation time increases. The choice of $l$ is thus a trade-off between speed and precision. We follow Ref.~\cite{ci} and set $l=3$. Just like coreness, the collective influence needs all the network information. Thus the total cost of information gathering is $Nc_\mathrm{info}$.

\subsection{Networks}

Ideally, the underlying network of our study should be as realistic as possible (given a pathogen). Our knowledge of the structure of contact networks is advancing, and there are some datasets available. We use the ones that record actual contacts between people and disregard those where contacts are inferred from interaction on social media, etc.~\cite{blasio}. To better understand how the size of the network, and higher-order structures, affect the performance of the algorithms, it is desirable to have models able to generate contact networks. We study one of the simplest such models---the configuration model---not because it is able to generate a network with very realistic structure, but because it enables us to compare the result to other studies, in particular analytical ones.

\subsubsection{Configuration model}

The input to the configuration model is a degree sequence, i.e.\  a sequence of desired degrees of the nodes of the network. Then the model proceeds by picking random pairs of nodes and adding an edge between them if their actual degrees are less than their desired degrees.  When all nodes  has their desired degree, the network has been constructed. The model does not enforce a simple graph (i.e.\ if there are already edges between a selected pair of nodes, one would still add another edge, and links from a vertex to itself are also allowed).  Since the empirical graphs in our study are simple graphs by construction, we convert the output of the configuration model to a simple graph by deleting multiple edges and self-loops. In the literature this construction is sometimes called the  erased configuration model~\cite{Hofstad}. 

Like many previous studies, we focus on networks with a power-law degree distribution, so the probability of a vertex having degree $k$ is proportional to $k^{-\gamma}$.  We truncate the degree distribution at $N^{1/(\gamma-1)}$.  Such truncation improves the precision of the estimated average values of the infection outbreak, and at the same time it preserves  the limiting degree distribution and the order of magnitude of the maximum degree.

The parameter values we use are: $\gamma = 2.5$ (as a typical value of empirical networks) and $N=625$, $1,250$, $2,500$, $5,000$, or $10,000$. We generate $100$ networks of each combination of parameter values.

\begin{table}
\caption{\label{tab:tab}Basic statistics of the data sets. $N$ is the number of individuals; $M$ is the number of links. $x$ is the connectance (fraction of vertex pairs that are links). $C$ is the clustering coefficient of the original network and $C'$ denotes the averaged values of random graphs with the same expected degree sequence as the original network~\cite{bayati}.}
\begin{tabular}{l|ccccc}
Data set & N & M & $x$ & C & C' \\ \hline
\textit{HIV} & 40 & 41 & 0.052 & 0.034 & 0.094 \\
\textit{Colorado Springs} & 324 & 345 & 0.004 & 0.026 & 0.029 \\
\textit{Iceland} & 75 & 114 &  0.041 &  0.16 & 0.20 \\
\textit{Prostitution} & 16,730 & 39,044 & 0.0002 & 0 & 0.010 \\
\textit{Conference} & 113 & 2,196 & 0.34 & 0.50 & 0.48 \\
\textit{Hospital} & 75 & 1,139 & 0.41 & 0.58 & 0.57 \\
\textit{School 1} & 236 & 5,901 & 0.21 & 0.43 & 0.28 \\
\textit{School 2} & 238 & 5,541 & 0.20 & 0.47 & 0.27\\
\end{tabular}
\end{table}

\begin{figure*}
\includegraphics[width=\textwidth]{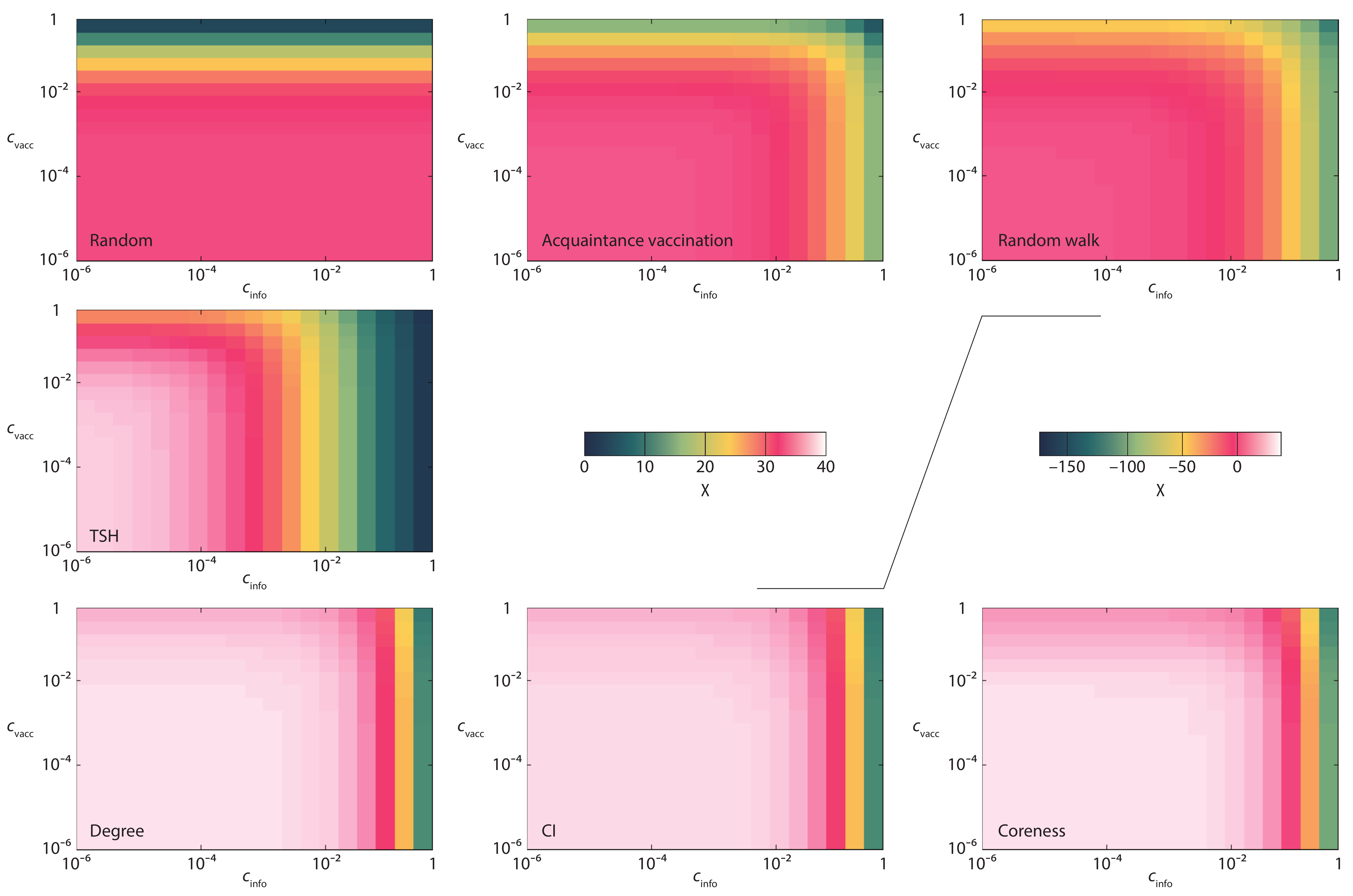}
\caption{{\bf An example of the cost efficiency of different vaccination strategies.} Here we use the \textit{Colorado Springs} network as a function of the costs of information retrieval and vaccination and $\beta=2$.}
\label{fig:opti_colorado_1}
\end{figure*}

\subsubsection{Empirical networks}

The first type of empirical networks that we use represent self-reported sexual contacts. Two of these data sets---we label them \textit{HIV} and \textit{Colorado Springs}---were gathered by so called contact tracing where individuals testing positive with HIV were required to report their recent contacts. \textit{HIV} data set is from the first study~\cite{hiv}, which used an observed contact network between HIV patients to argue that HIV is a sexually transmitted disease. \textit{Colorado Springs} is a larger and  more recent contact-tracing data set based on patients  from its namesake city in Colorado, USA~\cite{colorado_springs}. Contact tracing does not follow contacts of uninfected individuals, indeed \textit{HIV} only includes positive cases while \textit{Colorado Springs} also includes uninfected individuals that had sex with HIV positive others.

We also use two networks of self-reported sexual contacts not related to contact tracing. One (\textit{Iceland}) comes from Icelandic men who have sex with men~\cite{iceland}. The other (\textit{Prostitution}) from a Brazilian web forum where sex buyers report their encounters with prostitutes~\cite{prostitution}.

The final type of empirical networks are so called proximity networks. In these, a link represent a pair of people being close to each other at some time. These data sets all come from the Sociopatterns project (sociopatterns.org) and were collected by radio-frequency identification sensors given to people in some specific social setting. Such sensors record a contact if two persons are within 1--1.5 m.  The social setting of one of these data sets is a conference~\cite{conference} (\textit{Conference}), another is a hospital~\cite{hospital} (\textit{Hospital}) and the final one is from a school (\textit{School 1} and \textit{2})~\cite{school}.

The original proximity data sets along with \textit{Prostitution} are time resolved. We construct static networks by aggregating all contacts. (Ideally these data sets should be analyzed as temporal networks~\cite{naoki_holme}---then one could get around the assumption that the past accurately predicts the future~\cite{holme_tempo_vacci,starnini2013immunization}. However, that is outside the scope of this paper.)

We list the basic statistics---sizes, sampling durations, etc.---of the data sets in Table~\ref{tab:tab}.

\section{Results}
\subsection{Numerical results}

We start by evaluating the vaccination protocols in some detail for the \textit{Colorado Springs} data set. Then we proceed to take a cruder look at all the data sets to see how network structure affects the results.

\subsubsection{A case study}

The \textit{Colorado Springs} network serves well as an example since it is of intermediate size in our collection and has typical features, such as a heterogeneous degree distribution. In this section we set $\beta=2$---once again choosing a modest value that is in the interesting range where disease can spread throughout the population. In Fig.~\ref{fig:opti_colorado_1}, we plot the optimal saved cost $N\chi_\mathrm{opt}$ as a function of the two parameters---the relative cost of information $c_\mathrm{info}$ and the relative cost of vaccination $c_\mathrm{vacc}$. The general pattern is quite trivial---the protocols needing most information (\textit{CI}, \textit{Degree} and \textit{Coreness}) are also the ones that depend most on $c_\mathrm{info}$, while \textit{Random}, that needs no information at all, depends only on $c_\mathrm{vacc}$. The three protocols using an amount of information depending on $f$ (\textit{Acquaintance}, \textit{Random walk} and \textit{TSH}) are affected by both $c_\mathrm{info}$ and $c_\mathrm{vacc}$. From the heat maps it is hard to see which protocol is the best (except, perhaps that \textit{Acquaintance} has the largest $\chi$ for large $c_\mathrm{info}$). This means that the efficiencies of the best-performing protocols are relatively similar.

\begin{figure*}
\includegraphics[width=0.65\textwidth]{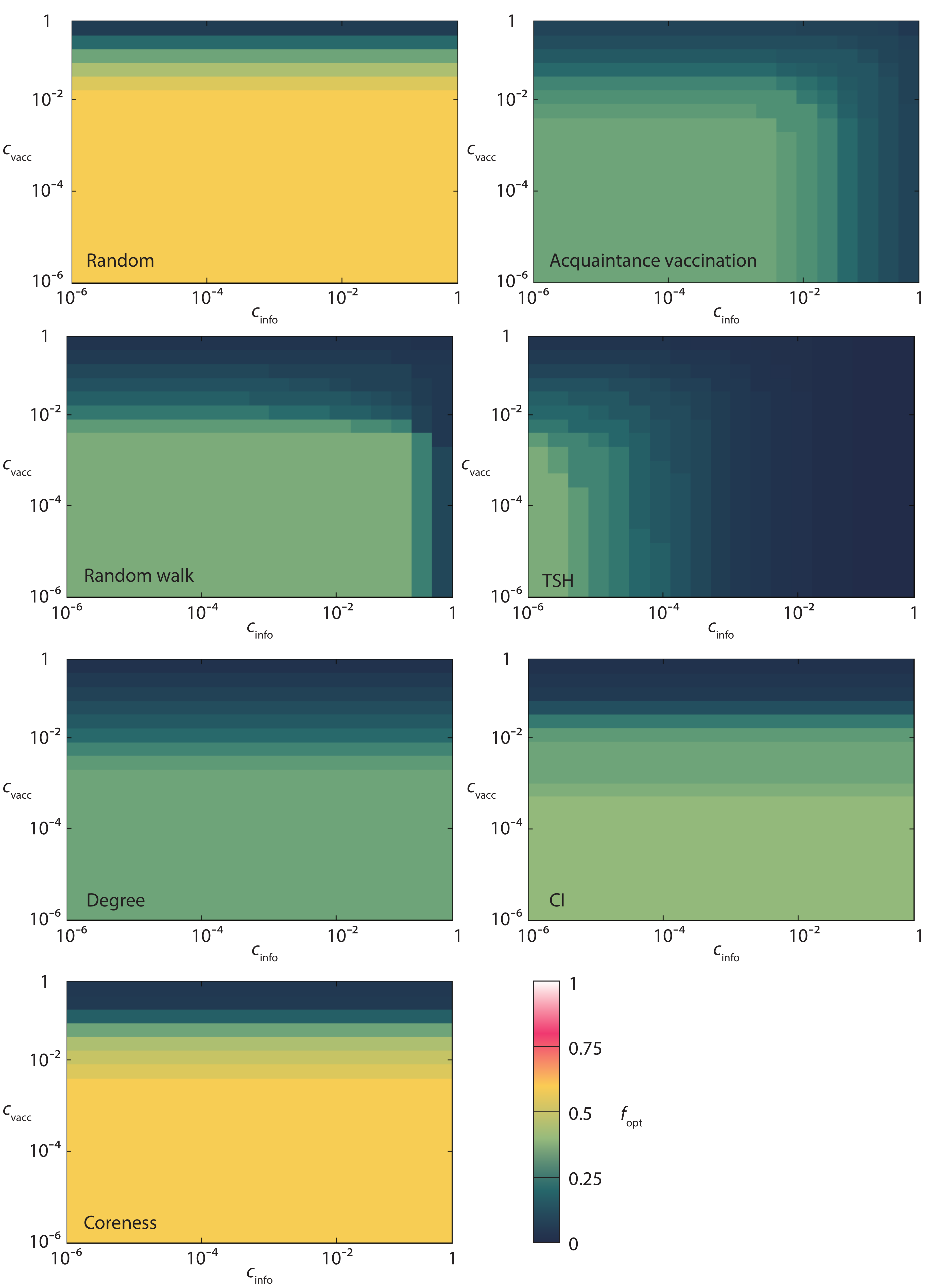}
\caption{{\bf The optimal fraction of vaccinated vertices for the same situation as in Fig.~\ref{fig:opti_colorado_1}.}
\label{fig:fracvac_colorado_1}}
\end{figure*}

\begin{figure}
\includegraphics[width=\columnwidth]{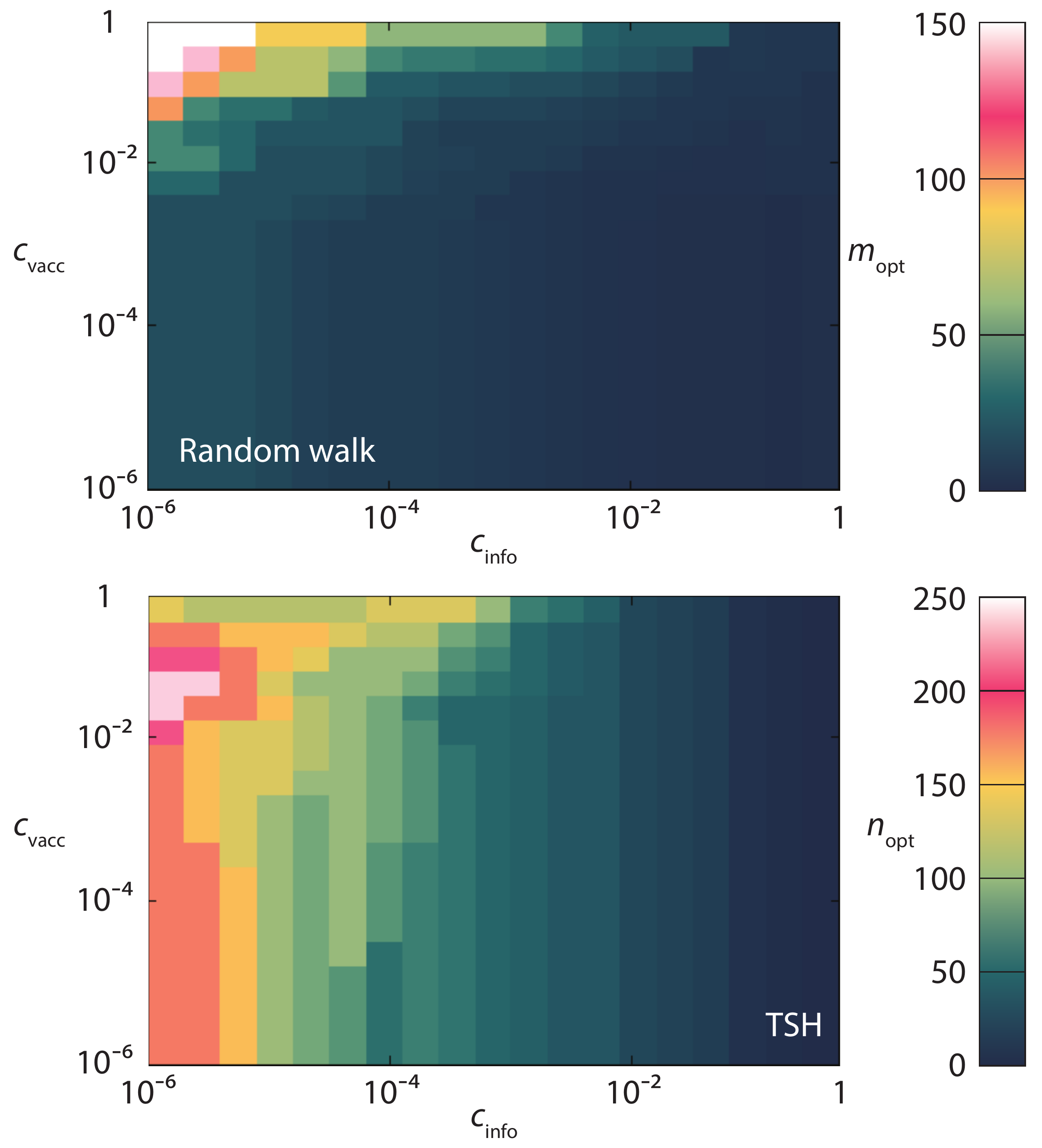}
\caption{{\bf The parameter values optimizing the \textit{Random walk} and \textit{TSH} strategies.} The plot shows the same network and as a function of the same parameters as in Fig.~\ref{fig:opti_colorado_1}.}
\label{fig:params_colorado_1}
\end{figure}

The performance of the protocols can be better understood by measuring the fraction of vertices $f_\mathrm{opt}$ needed to be vaccinated to optimize the total costs. See Fig.~\ref{fig:fracvac_colorado_1}. The protocols where the information costs do not depend on $f$ obviously have no $c_\mathrm{info}$ dependence. For the other ones---\textit{Acquaintance}, \textit{Random walk} and \textit{TSH}---$f_\mathrm{opt}$ decreases with both $c_\mathrm{vacc}$ and $c_\mathrm{info}$. Hence, more information does make these protocols more accurate. This can be seen even more clearly in Fig.~\ref{fig:params_colorado_1} where we set $f=f_\mathrm{opt}$ and study the optimal parameter values ($m_\mathrm{opt}$ and $n_\mathrm{opt}$) of the \textit{Random walk} and \textit{TSH} protocols. Both the protocols naturally have larger values of their parameters the cheaper the information is. For \textit{Random walk} the optimal $m$-value is largest when $c_\mathrm{info}$ is as small and $c_\mathrm{vacc}$ as large as possible. Large $c_\mathrm{vacc}$ gives small optimal $f$ (see Fig.~\ref{fig:fracvac_colorado_1}) which lowers the cost needed for gathering information. For large $c_\mathrm{vacc}$ and small $c_\mathrm{info}$ the relative cost for information gathering is thus so small that the rather small marginal benefit of longer random walks is still affordable.

For the \textit{TSH} protocol the largest parameter value is at an intermediate value of $c_\mathrm{vacc}$ (still $c_\mathrm{info}$ is as small as possible). One can understand the increase of the parameter value with $c_\mathrm{vacc}$ in a similar way as for \textit{Random walk}. The eventual decrease, for $c_\mathrm{vacc}\approx 0.1$, as well as other non-monotonicities in the plot, can be related to how $\Omega'$ responds to changing  $f_\mathrm{opt}$.

\begin{figure*}
\includegraphics[width=\textwidth]{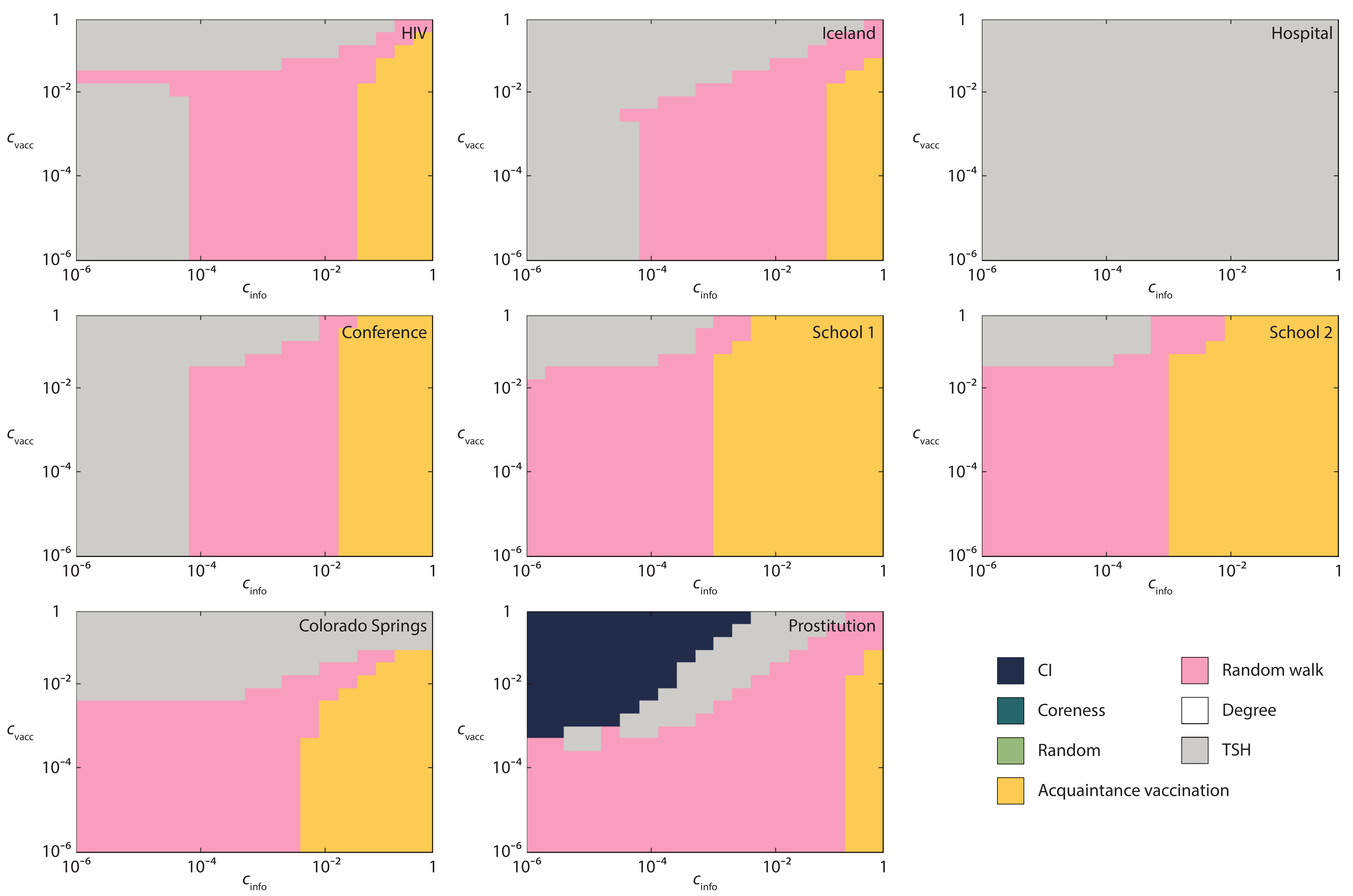}
\caption{{\bf The most cost efficient vaccination strategies for empirical networks as a function of the costs of information retrieval and vaccination.} In this figure $\beta=2$.}
\label{fig:cmpr1}
\end{figure*}

\begin{figure*}
\includegraphics[width=1\textwidth]{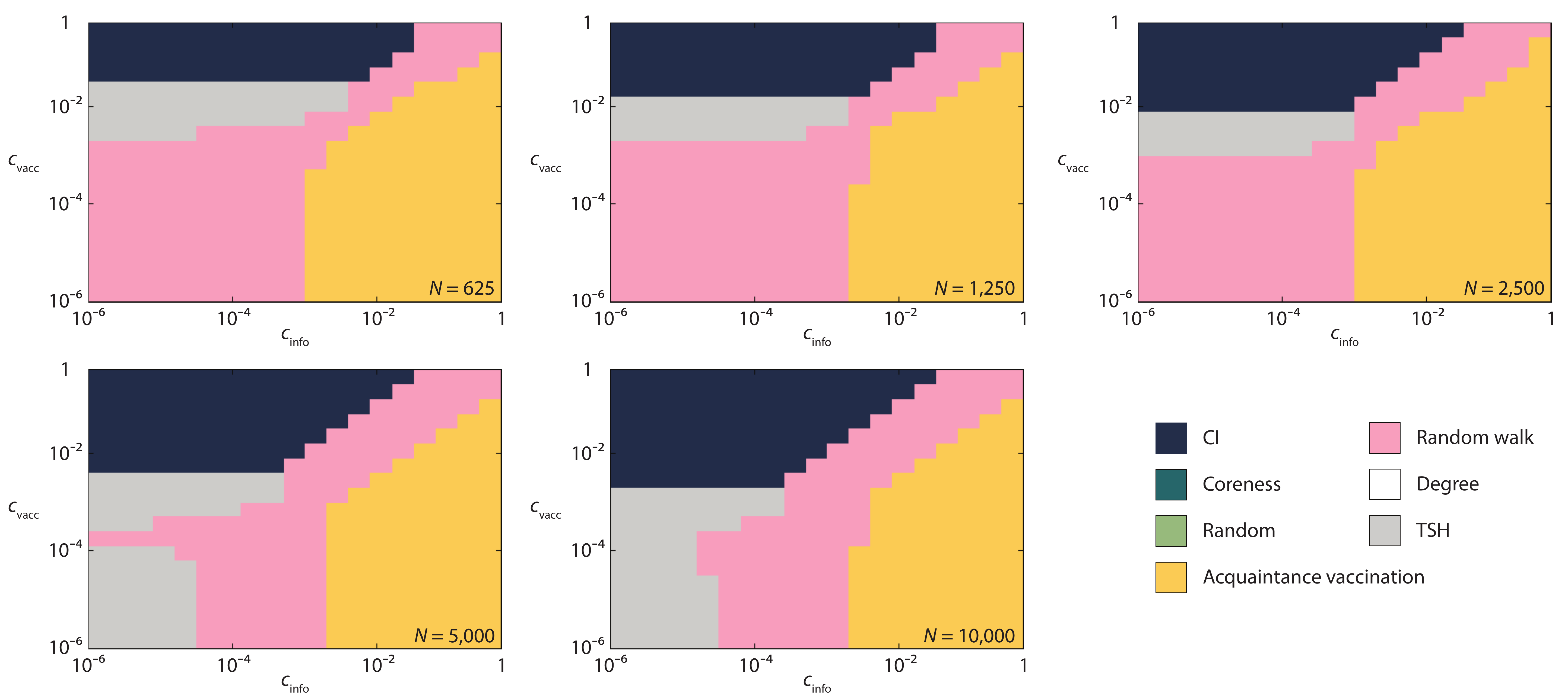}
\caption{{\bf The most cost efficient vaccination strategies for the configuration model with a power-law degree distribution as a function of the costs of information retrieval and vaccination.} Like in Fig.~\ref{fig:cmpr1}, $\beta=2$.}
\label{fig:cmpr_arti1}
\end{figure*}

\subsubsection{Network-structural effects}

The picture painted in the previous section remains roughly true for other data sets and $\beta$ values. In this section, we go directly to our main question of what the most cost effective vaccination protocol is. Figure~\ref{fig:cmpr1} shows the results for $\beta=2$. The corresponding figure for the other $\beta$-values we study can be found in the Supplementary material. From these figures, the conclusions are roughly the same, but for small $\beta$, i.e.\ small outbreak sizes, the results are affected by noise (so the regions are not that clear cut).

For most of the data sets, \textit{Acquaintance} vaccination is the most efficient protocol for relatively large information costs, \textit{TSH} is the most efficient for low $c_\mathrm{info}$ and large $c_\mathrm{vacc}$, while \textit{Random walk} is the most efficient for the rest of the parameter space. One exception is the \textit{Prostitution}---the largest and sparsest network---where \textit{CI} is the most cost effective (despite the fact it requires global knowledge of the network structure). This network also has zero clustering coefficient---i.e.\ no triangles (because only heterosexual contacts are recorded). Still, the size and sparsity seem like more fundamental differences to the other networks (cf.\ Ref.~\cite{holme_tempdis}). To understand the role of clustering one could perform the same study on model networks where the clustering can be controlled. The densest network, \textit{Hospital}, is also different in the respect that \textit{TSH} performs best for the entire parameter space. \textit{Random} is never the most efficient, meaning that there are network structures that can be exploited for all data sets and parameter values. \textit{Coreness} and \textit{Degree} does not perform best under any circumstance.

In addition to the empirical contact networks, we also study scale-free networks of different sizes. See Fig.~\ref{fig:cmpr_arti1}. These networks behave slightly differently from the empirical networks with \textit{CI} dominating the large-$c_\mathrm{vacc}$ small-$c_\mathrm{info}$ region, \textit{Acquaintance} dominating the small-$c_\mathrm{vacc}$ large-$c_\mathrm{info}$ region, \textit{Random walk} being the best for the region of intermediate $c_\mathrm{vacc}$ and $c_\mathrm{info}$, and \textit{TSH}  being the best protocol for some low $c_\mathrm{info}$ values and intermediate $c_\mathrm{vacc}$ values.

\subsubsection{Analysis of the optimal $f$}

  In this section our goal is to understand regularities behind the numerical results. The exact analysis is available only for asymptotic behavior of \textit{Random}, \textit{Acquaintance} and \textit{Degree} strategies in a configuration model, see e.g.~\cite{Britton2007graphs,Lelarge2009efficient_vacc}, but the analytical expressions are cumbersome, and do not provide sufficient qualitative insights. The results on other vaccination strategies are currently not available. Therefore, we resort to heuristic arguments, that are based on the exact results in the literature.

Dividing both parts of Eq.~(\ref{eq:chi}) by $N$, we write
\begin{equation}
\label{eq:chi1}
\chi(f) = \Delta(f) -fc_\mathrm{vacc} - [n(f,G)/N]c_\mathrm{info},
\end{equation}
where 
\begin{equation}\Delta(f)=(\Omega -\Omega'(f))/N\end{equation}
 is the fraction of the population that have avoided the disease due to vaccination. For any vaccination strategy, $\Delta(f)$, obviously, increases in $f$. Furthermore, remember that $f_c$ is the fraction of the population that needs to be vaccinated in order to prevent a global outbreak. In other words, if $f\ge f_c$, then $\Omega'(f)$ is negligible compared to $N$, so $\Delta(f)\approx\Omega/N$. Moreover, one expects that for small $\epsilon>0$, the additional gain $\Delta(f+\epsilon)-\Delta(f)$ decreases to zero when $f$ approaches $f_c$. (This is closely related to subadditivity of spreading processes, which is used, for example, in solving influence maximization problems~\cite{Kempe2003maximizing}.) 

\begin{figure}[h]\includegraphics[width=0.8\columnwidth]{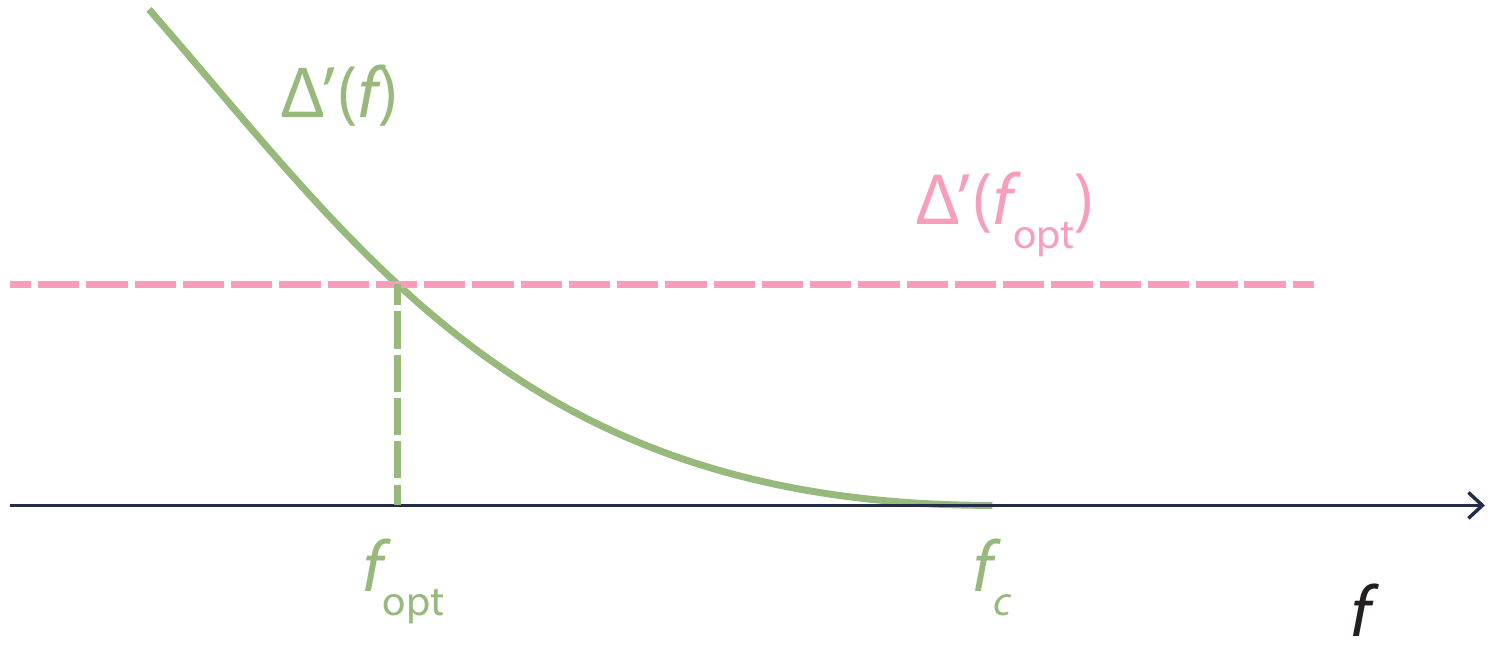}
\caption{{\bf Schematic representation of $\Delta'(f)$.} The value $\Delta'(f_\mathrm{opt})$ is given by Eq.~(\ref{eq:derivative}).}
\label{fig:delta_prime}
\end{figure}

Following a widely used approach in epidemiology and network science, consider a continuous version of Eq.~\eqref{eq:chi1}, where all functions of $f\in [0,1]$ are differentiable and all vanishing terms are neglected. (We note that proving formally that the process converges to its continuous representation as $N\to\infty$ is a challenging mathematical problem, however, it is common to analyze the continuous version  on its own rights.) In the continuous version of the system, our observations above can be summarized as follows: (i) $\Delta'(f)>0$ for $f<f_c$; (ii) $\Delta(f)=\Omega/N$ and $\Delta'(f)=0$ for $f\ge f_c$; (iii) $\Delta'(f)\to 0$ when $f\to f_c$. This behavior of $\Delta'(f)$ is schematically depicted in Fig.~\ref{fig:delta_prime}.

We proceed with analyzing the optimal fraction of vaccinated individuals. 
 The maximal gain in Eq.~(\ref{eq:chi1}) is achieved at $f=f_\mathrm{opt}$, which is a solution of
\begin{equation}
\label{eq:derivative}
\Delta'(f)=c_\mathrm{vacc} + [n'(f,G)/N]c_\mathrm{info}.
\end{equation}
 Since $n(f,G)$ is non-decreasing in $f$, it follows that  $\Delta'(f_\mathrm{opt})>0$, so $f_\mathrm{opt}<f_c$. One can now view $f_c$ as an indicator of the effectiveness of a vaccination strategy in preventing the epidemics. Indeed, when $f_c$ is small, then the global outbreak is prevented by vaccinating only a small fraction of individuals. Consequently,  one expects that $f_\mathrm{opt}$ is smaller for more effective strategies. Another indication for smaller $f_\mathrm{opt}$ is a higher value of the right-hand side of Eq.~(\ref{eq:derivative}), as can be easily seen in Fig.~\ref{fig:delta_prime}, where this value is represented by the dashed line.

We will now compare $f_\mathrm{opt}$ for different vaccination strategies.  


{\it Random} ({\it R}) vaccination strategy is most understood in the literature. Assume that the underlying graph is a configuration model. If the degree distribution has a finite variance, then $f_c^R$ can be obtained directly from Eq.~(3.5) in Ref.~\cite{Britton2007graphs} by equating the reproduction number to its critical value~1. Specifically, we have:
\begin{equation}f^R_c= \max\left\{1-\frac{\langle k \rangle}{\langle k^2\rangle-\langle k\rangle}\cdot \frac{\beta+\nu}{\beta},0\right\},\end{equation}
and the value is positive if the global outbreak occurs when no vaccination takes place. When the variance is infinite, as in our case $\gamma=2.5$, then $f^R_c=1$, so the global outbreak cannot be prevented by the random vaccination. 

Applying Eq.~(\ref{eq:derivative}) we obtain  that $f^R_\mathrm{opt}$ satisfies
\begin{equation}
\label{eq:fR}
\Delta'(f)=c_\mathrm{vacc}.\end{equation}
When $f^R_c=1$, one expects that $f^R_\mathrm{opt}$ is quite large for small $c_\mathrm{vacc}$, and it decreases when $c_\mathrm{vacc}$ becomes larger (see Fig.~\ref{fig:delta_prime}).  This is indeed the case in our case study in Fig.~\ref{fig:fracvac_colorado_1}. We can also explain the modest gain in Fig.~\ref{fig:cmpr1} by relatively slow growth of $\Delta(f)$.


  For the {\it Acquaintance} ({\it A}) strategy, in the configuration model, $f^A_{c}\in (0,1)$ can be computed using Theorem~3.3 of Ref.~\cite{Britton2007graphs},  as long as the reproduction number in Eq.~(3.13) in Ref.~\cite{Britton2007graphs} is smaller than one. The optimal fraction of vaccinated individuals $f^A_\mathrm{opt}$ satisfies
\begin{equation}\Delta'(f) = c_\mathrm{vacc}+c_\mathrm{info}.\end{equation}
Compared to the {\it Random} strategy, the right-hand side has an extra positive term. 
Moreover, for the same epidemic on the same graph, it holds that $f^A_{c}< f^R_{c}$ (except when $f^A_{c}=f^R_{c}=1$), see Ref.~\cite{Britton2007graphs}. { Hence, using Fig.~\ref{fig:delta_prime}, we deduce that $f^A_\mathrm{opt}$ should be considerably smaller than $f^R_\mathrm{opt}$.  We see that this is indeed the case in Fig.~\ref{fig:fracvac_colorado_1}. 

The cost efficiency of {\it Acquaintance} and {\it Random} strategies is harder to compare because {\it Acquaintance} targets high-degree nodes while {\it Random} does not, but on the other hand, {\it Random} has no information costs. To take extreme examples, {\it Random} will yield higher gain on a regular graph, while {\it Acquaintance}---on a star graph. In the case study in Fig.~\ref{fig:cmpr1}, we see that the gain for the {\it Acquaintance} strategy is similar to the one for the {\it Random} strategy, while in other data sets {\it Acquaintance} outperforms other protocols especially when information costs are high, see Fig.~\ref{fig:cmpr1}.


{\it Degree} ({\it D}), {\it Coreness} ({\it C}) and  {\it CI} strategies must be most effective in the configuration model because they target the nodes that have the highest potential for spreading the infection. A formula for the average outbreak size in the configuration model when nodes of degree $s$ are removed with given probability is given in Ref.~\cite{Lelarge2009efficient_vacc}, but these results do not directly apply when fraction $f$ of highest degree nodes is removed. 

The fraction $f_\mathrm{opt}$ for these strategies satisfies the same equation (\ref{eq:fR}) as $f^R_\mathrm{opt}$, that is, $\Delta'(f) = c_\mathrm{vacc}$.
 A comparison between $f_{\mathrm{opt}}^D$ and $f_{\mathrm{opt}}^A$ may go both ways, as is easily illustrated by Fig.~\ref{fig:delta_prime}. On one hand, one expects that $f^\mathit{D}_c<f^A_c$, because both strategies target high degree nodes, only {\it Degree} identifies them precisely while {\it Acquaintance} is just a heuristic. On the other hand, $\Delta'(f_\mathrm{opt})$ is smaller for the {\it Degree} than for the {\it Acquaintance} strategy. Same argument applies to {\it Coreness} and {\it CI}, however, these protocols do not per se target nodes of large degrees, so depending on a network, $f^\mathit{C}_c$ and $f^\mathit{CI}_c$ might be smaller or larger than $f^A_c$.

In the case study in Fig.~\ref{fig:fracvac_colorado_1} we obtain $f^\mathit{D}_\mathrm{opt}<f^A_\mathrm{opt}$ but 
$f^\mathit{C}_\mathrm{opt}, f^\mathit{CI}_\mathrm{opt}>f^A_\mathrm{opt}$. Very large value of $f_\mathrm{opt}$, especially for {\it Coreness} in Fig.~\ref{fig:fracvac_colorado_1} signals that these strategies are in fact inefficient for the {\it Colorado Springs} case study. In Fig.~\ref{fig:cmpr1}, for the same case study we observe that {\it Degree}, {\it Coreness} and  {\it CI} have a very small gain. The efficiency of {\it CI} on configuration model (Fig.~\ref{fig:cmpr_arti1}) and on the {\it Prostitution} data set in Fig.~\ref{fig:cmpr1}, for similar values of the parameters, is an interesting finding that deserves further research. Possible explanation can be in a small number of triangles---the feature that the {\it Prostitution} data set and configuration model share.


Finally, consider {\it Random-walk (RW)} and {\it TSH} strategies. Since these protocols target nodes with large degrees, but do not identify them precisely, one expects that the {\it Degree} protocol is more effective in preventing a global outbreak, but not by much. Therefore, $f_c^\mathit{RW}$ and $f_c^\mathit{TSH}$ should be slightly larger than  $f^\mathit{Degree}_c$. The optimal value $f_\mathrm{opt}$ satisfies
\begin{equation}\Delta'(f) = c_\mathrm{vacc}+n'(f,G)c_\mathrm{info}/N.\end{equation}
For large enough $N$, we expect the last term above to be small, so $\Delta'(f_\mathrm{opt})$ is close to the one of the {\it Degree} protocol. Invoking Fig.~\ref{fig:delta_prime}, we expect that $f^\mathit{RW}_\mathrm{opt}$ and $f^\mathit{TSH}_\mathrm{opt}$ are close to $f^\mathit{Degree}_\mathrm{opt}$, especially when $c_\mathrm{info}$ is small, and they decrease when $c_\mathrm{info}$ increases. These are exactly the results in Fig.~\ref{fig:fracvac_colorado_1}. The net gain  of {\it Random walk} and {\it TSH} should be considerably higher than that of the {\it Degree} strategy when $c_\mathrm{info}$ is large enough. Indeed, we observe that the {\it Degree} strategy never yields the largest gain.

The comparison of {\it Random walk} and {\it TSH} to the {\it Acquaintance} strategy is trickier since the latter also targets high degree nodes but at lower costs. On the other hand, the accuracy of {\it Random walk} and {\it TSH} is  higher. The comparison between the three randomized strategies---{\it Acquaintance}, {\it Random walk}, and {\it TSH}---thus depends on the interplay between accurate targeting and information costs. This explains that the {\it Acquaintance} sometimes performs better than {\it Random walk} and {\it TSH}.

\section{Discussion}

 We have discussed how to make theoretical studies of targeted vaccination more practically useful for decision makers. Instead of evaluating vaccination protocols for some scenario about what is known about the network, we evaluate methods based as a cost-benefit problem. From this starting point, we have evaluated the cost efficiency of seven network-based vaccination methods. There is not one universally best method. Rather, depending on the network structure and relative vaccination and information costs, the best method (at least for the network and parameters we explore) seem to be one of four---\textit{Acquaintance}, \textit{TSH}, \textit{CI} and \textit{Random walk}. We make this point both by analytical calculations and simulations.

\textit{Acquaintance} vaccination is almost always the most efficient for small $c_\mathrm{vacc}$ and large $c_\mathrm{info}$. It is the protocol that uses second least network information after \textit{Random}. For very large $c_\mathrm{info}$, \textit{Random} will trivially be the most efficient (keep in mind that $c_\mathrm{info}$ can, in principle, be larger than one), but we never observe this. \textit{TSH} dominates the region of large $c_\mathrm{vacc}$ and small $c_\mathrm{info}$, for denser networks (for sparser networks \textit{CI} could also be most efficient). \textit{Random walk} dominates intermediate values of $c_\mathrm{vacc}$ and $c_\mathrm{info}$. It is hard to speculate why, but it is probably related to the fact that the optimal parameter values of \textit{TSH} change quickly for small $c_\mathrm{info}$ (Fig.~\ref{fig:params_colorado_1}). In other words, one can tune the performance-to-cost balance more, meaning that this protocol is more adaptable than \textit{Random walk} in this region. \textit{CI} performs well for sparse networks with few triangles, especially in the region of large $c_\mathrm{vacc}$ and small $c_\mathrm{info}$. \textit{Degree} is never most efficient, meaning that vaccinating exactly in order of degree is not so important that it is worth obtaining all the network information. Furthermore, \textit{Coreness} is also never most efficient, supporting Refs.~\cite{holme_pcb} and \cite{ci} (but disagreeing with Ref.~\cite{core}).

In practical applications, one would in principle need to know the parameters, both for the SIR model and to calculate the cost~\cite{wang_etal_statphys}. For e.g.\ sexually transmitted diseases, this is not impossible. If one, would base a pilot HIV pre-exposure prophylaxis campaign on mapping a sexual network like Ref.~\cite{iceland} (which, in addition to the network itself, could give the contact rates), then one could assume a per-contact transmission probability of 1--2\%~\cite{catie}. Furthermore, the societal cost for a positive HIV case is well understood~\cite{hivsoc}. With these parameters at hand, it should be possible to narrow down the protocols to one or two.

 To proceed towards increasing realism and applicability, one would also need to take social mechanisms into account. Parallel to the targeted immunization problem, there is an emergent field studying vaccination as a social-psychological problem. One issue being that for voluntary vaccination it is irrational to become vaccinated if almost everyone else is vaccinated (the diseases would not spread anyway, and there are side-effects and discomfort associated with being vaccinated). Conversely, it is irrational not to vaccinate if almost nobody is vaccinate, leading to a typical game theoretical dilemma~\cite{zhixi_rev}. Another issue in this direction discusses how the awareness of a disease spreading affect the contact networks, and subsequently the spreading dynamics~\cite{funk}. Or how vaccination and awareness diffusion can create synergistic effects~\cite{vaccine_adaptive}. Other papers study how social influence affects the decision to vaccinate ones children (e.g.\ Ref.~\cite{brunson}). To make theoretical vaccination studies fully realistic and most useful to decision makers, one would need combine such social aspects with the cost-benefit approach of this paper.

\begin{acknowledgments}
We are grateful to Tom Britton and Maria Deijfen for the very useful discussion.
\end{acknowledgments}

\end{document}